\documentclass[oldversion]{aa} 
\usepackage{graphicx}
\usepackage{txfonts}
\usepackage{natbib}
\bibpunct{(}{)}{;}{a}{}{,}

\pdfoutput=0

\begin{document}
 
\title{On the magnetic topology of partially and fully
  convective stars}

\author{A.~Reiners
  \inst{1}\fnmsep\thanks{Emmy Noether Fellow}
 \and
 G.~Basri\inst{2}
}


\institute{
  Universit\"at G\"ottingen, Institut f\"ur Astrophysik, Friedrich-Hund-Platz 1, D-37077 G\"ottingen, Germany\\
  \email{Ansgar.Reiners@phys.uni-goettingen.de}
  \and
  Astronomy Department, University of California, Berkeley, CA, 94720, USA\\
  \email{basri@berkeley.edu}
}

\date{Received ... / Accepted ...}


\abstract {We compare the amount of magnetic flux measured in Stokes~V
  and Stokes~I in a sample of early- and mid-M stars around the
  boundary to full convection ($\sim$M3.5). Early-M stars possess a
  radiative core, mid-M stars are fully convective.  While Stokes~V is
  sensitive to the net polarity of magnetic flux arising mainly from
  large-scale configurations, Stokes~I measurements can see the total
  mean flux.  We find that in early-M dwarfs, only $\sim 6$\,\% of the
  total magnetic flux is detected in Stokes~V. This ratio is more than
  twice as large, $\sim 14$\,\%, in fully convective mid-M dwarfs. The
  bulk of the magnetic flux on M-dwarfs is not seen in Stokes~V. This
  is presumably because magnetic flux is mainly stored in small scale
  components. There is also more to learn about the effect of the
  weak-field approximation on the accuracy of strong field detections.
  In our limited sample, we see evidence for a change in magnetic
  topology at the boundary to full convection.  Fully convective stars
  store a 2--3 times higher fraction of their flux in fields visible
  to Stokes~V.  We estimate the total magnetic energy detected in
  Stokes~I and compare it to results from Stokes~V.  We find that in
  early-M dwarfs only $\sim$0.5\,\% of the total magnetic energy is
  detected in Stokes~V while this fraction is $\sim$2.5\,\% in mid-M
  dwarfs.  }

\keywords{stars: late-type -- stars: magnetic fields -- stars:
  activity}

\maketitle
%

\section{Introduction}
\label{sect:Introduction}

Magnetic fields are ubiquitous in cool stars. There is growing
evidence that their total strength is mainly a question of rotation,
and that the efficiency of magnetic field generation follows similar
rules in solar-type stars and much cooler objects including planets
\citep{Christensen08}. While magnetic field generation in solar-type
stars is probably most efficient near the interface layer between the
radiative core and the convective envelope
\citep[e.g.,][]{Charbonneau05}, the dynamo process in fully convective
stars must be different. The transition from partially to fully
convective interiors happens around a mass of 0.35\,M\,$_\odot$. In
lower-mass stars, mean magnetic flux does not significantly differ
from partially convective stars \citep{Reiners07}, although they
probably operate a different type of dynamo \citep[e.g.,][]{Durney93}.

The solar magnetic field is predominantly axisymmetric and dipolar
\citep[see, e.g.,][]{Ossendrijver03}, which we know from direct
imaging of the Sun. For other stars, no direct images are available
and we have to rely on indirect methods to investigate magnetic
topologies. Magnetic fields are usually measured in different Stokes
components via the Zeeman effect. Two main methods can be
distinguished: 1) Zeeman splitting in Stokes~I measures the total mean
magnetic flux on the star, and 2) Stokes~V measures effective magnetic
polarization. In order to completely characterize the magnetic
topology of a star, in principle all four Stokes components are
necessary \citep[e.g.,][]{Kochukhov02}, but this is observationally
extremely difficult to achieve. The method of Zeeman Doppler Imaging
in Stokes~V has been very successfull during the last years providing
the first information on the magnetic structure on low-mass stars
\citep[e.g.,][and references therein]{Donati08, Morin08}. Using
Stokes~V, however, implies the problem that only the net polarization
is visible and that probably much of the magnetic flux cancels out and
is unobservable. The observation of dense time series helps resolving
smaller structures but cannot entirely solve that problem. Because
Stokes~I is sensitive to the total magnetic flux, and Stokes~V is
sensitive to the net (large-scale) polarization, a comparison of both
yields information about the amount of small scale flux that is
invisible to Stokes~V. The first question we address here is how much
magnetic flux escapes detection in Stokes~V Doppler maps.

Recently, \citet{Morin08} and \citet{Donati08} used magnetic maps from
Stokes~V to investigate magnetic topologies around the boundary to
full convection. They found that the fraction of axisymmetric fields
and the low-$l$ dipole modes are larger in fully convective stars than
in partially convective ones. Thus, the magnetic flux visible to
Stokes~V is more organized in fully convective stars. In this paper,
we ask the question whether a higher degree of organisation in fully
convective stars applies to the total magnetic flux, or to the flux
visible in Stokes~V only.

\section{Data}

Most of the data used for our analysis are taken from the literature.
Magnetic flux measurements from Stokes~V are taken from
\citet{Donati08} for the early-M dwarfs and from \citet{Morin08} for
mid-M dwarfs. These authors also provide rotational period and Rossby
numbers for the sample stars \citep[the latter taken
from][]{Kiraga07}.

Stokes~I measurements of magnetic flux are also taken from the
literature, and we present two new measurements in this work. The
magnetic flux of EV~Lac (Gl~873) was measured by \cite{JKV00} using a
detailed model of Zeeman splitting in an atomic line. This result was
used by \citet{Reiners07} to calibrate measurements of magnetic flux
using molecular FeH as a tracer. Our values of $Bf$ for AD~Leo
(Gl~388) and YZ~Cmi (Gl~285) are taken from \citet{Reiners07}.  For
DT~Vir (Gl~494A) we found a magnetic flux measurement in
\citet{Saar96} using atomic lines ($B=3.0$\,kG, $f=50$\,\%). Because
magnetic field measurements may be affected by the choice of
absorption lines used for analysis, we apply a correction factor to
the $Bf$ value of DT~Vir based on comparison between other $Bf$
measurements in \citet{Reiners07} and \citet{Saar96}; the four stars
AD~Leo, YZ~Cmi, EV~Lac, and Gl~729 were investigated in both works.
The magnetic flux $Bf$ reported by \citet{Reiners07} is systematically
higher than the values reported by \citet{Saar96}. Specifically, the
results for YZ~Cmi, EV~Lac, and Gl~729 are consistent if the filling
factor $f$ in \citet{Saar96} is assumed to be one. In other words, $B$
in \citet{Saar96} has roughly the same value as $Bf$ in
\citet{Reiners07}. Thus, we use $\langle B_I \rangle = 3.0$\,kG for
DT~Vir, which brings all values of $\langle B_I \rangle$ used here on
a consistent scale. We note that an uncertainty of a factor of 2 in
$\langle B_I \rangle$ would not influence the results of this paper.

We present additional magnetic flux measurements of two other stars in
this work.  Data of Gl~182 ($J$\,=\,7.1) were obtained with HIRES at
the W.M.~Keck observatory in the same way as reported in
\citet{Reiners07}. We used a slit-width of 1.15\arcsec\ yielding a
spectral resolving power of $R \approx 31,000$. The 600\,s exposure
has a SNR of about 120 at 9930\,\AA. A spectrum of CE~Boo (Gl~569A,
$J$\,=\,6.6) was obtained at the Hobby-Eberly-Telescope using HRS
centered at 8991\,\AA. The 2\arcsec\ fibre was used providing a
spectral resolving power of $R \approx 60,000$. After 300\,s a SNR of
40 was reached in the FeH band.

\section{New $Bf$ measurements}

\begin{figure*}
  \parbox{.48\hsize}{\center \resizebox{.9\hsize}{!}{\includegraphics{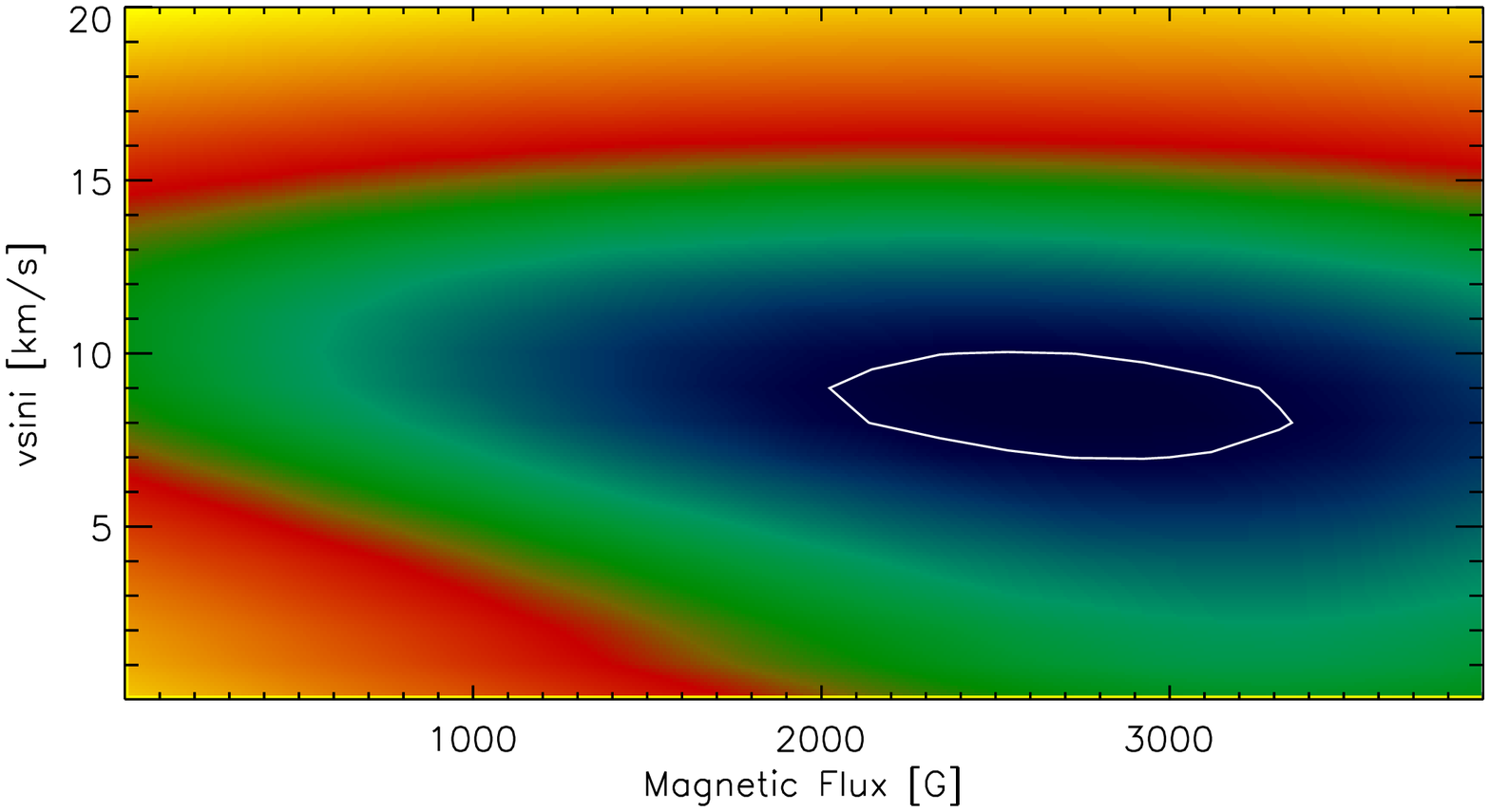}}}
  \parbox{.48\hsize}{\center \resizebox{.9\hsize}{!}{\includegraphics{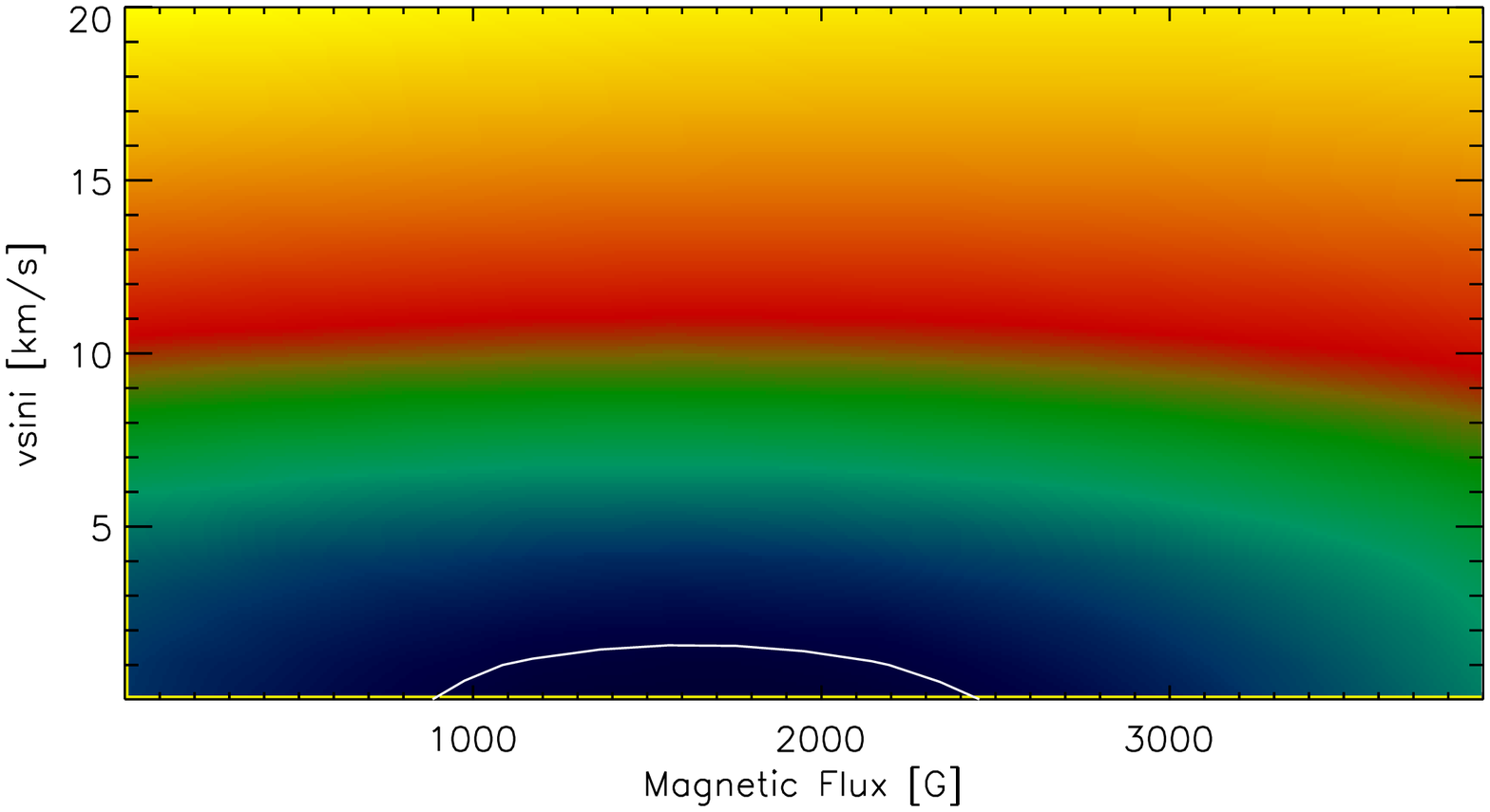}}}
  \caption{$\chi^2$-landscapes of our model to the data of Gl~182
    (left) and Gl~569A (right) as a function of projected rotation
    velocity $v\,\sin{i}$ and mean magnetic flux, $Bf$. Different
    colours indicated different values of $\chi^2$; blue dark colours
    mark areas with low $\chi^2$ (good fit), red and yellow are
    regions of high $\chi^2$ (bad fit). The white contour is
    sorrounding the region where $\chi^2 < \chi^2_{\rm min} + 9$,
    i.e., the 3-$\sigma$ region.\label{fig:stars} }
\end{figure*}

\subsection{Method}

The method we employ to measure the magnetic flux in Gl~182 and CE~Boo
was introduced in \citet{Reiners06} and after that applied to several
stars \citep[e.g., ][]{Reiners07, Reiners08}. Here, we give a brief
overview of the method and refer to the literature for a more detailed
description.

The absorption band of FeH contains a forest of strong, well isolated
lines of which some are sensitive to the Zeeman effect while others
are not. In a relatively small spectral range, we find spectral lines
of the same ro-vibrational transition that react differently to the
presence of a magnetic field. The direct simulation of the Zeeman
effect in FeH calculating polarized radiative transfer is still
hampered by the lack of Land\'e factors. Instead, we choose a more
empirical approach: We observed two spectra of early-M dwarfs for
which magnetic field measurements from atomic absorption lines exist.
One of them shows no signs of magnetic fields (and no activity;
GJ~1002), for the other a total magnetic flux of $Bf \sim 3.9$\,kG was
measured \citep[Gl~873,][]{JKV00}. We apply an optical-depth scaling
to the two reference spectra so that the strength of the FeH band
matches the strength of FeH absorption in the target star. The shape
of magnetically insensitive lines is used to fix the rotational
velocity, and magnetically sensitive lines to adjust for the magnetic
flux. This is done by interpolating the two template spectra (between
zero magnetic flux and $Bf =3.9$\,kG) in order to achieve the best fit
to the target spectrum.

\subsection{Magnetic flux in Gl~182 and CE~Boo}
\label{sect:Bfmeasurement}

The results of our fitting procedure are shown in
Fig.\,\ref{fig:stars}, where the goodness of fit in terms of $\chi^2$
is shown as a function of $Bf$ and $v\,\sin{i}$. The white line
surrounds the region of $\chi^2 < \chi^2_{\rm min} + 9$, i.e., the
3$\sigma$ limit \citep[see][for more details]{Reiners08}. In both
cases, the minimum value of the reduced $\chi^2$ is $\chi^2_{\rm min,
  \nu}~\approx~1$.  The formal results of our fitting procedure are,
for Gl~182, $v\,\sin{i} = 9 \pm 2$\,km\,s$^{-1}$, $Bf =
2730\pm600$\,G, and for CE~Boo, $v\,\sin{i} < 3$\,km\,s$^{-1}$, $Bf =
1750\pm800$\,G ($3\sigma$).

\section{Total flux and large-scale flux}

\subsection{Magnetic flux in Stokes V and I}

\begin{table*}
  \caption{\label{tab:results} Properties of the sample stars. Rotation period and results from Stokes~V measurements are from \citet{Donati08} and \citet{Morin08}. Rossby numbers are from \citet{Kiraga07}. Sources of Stokes~I measurements are given in the Table. The total magnetic energy is estimated from total magnetic flux using a scaling that involves a factor $f$. We use $f = 1.25$ (see text).} 
  \begin{tabular}{llccccccccc}
    \hline
    \hline 
    \noalign{\smallskip}
    Name & other & Mass & $P_{\rm rot}$ & $Ro$ & $\langle B_{V}\rangle $ & $\langle B_{I}\rangle $ & $\frac{\langle B_V\rangle }{\langle B_I\rangle }$ & $\langle B_V^2 \rangle $ & $ f \cdot \langle B_I \rangle ^2 $ &  $\frac{\langle B_V^2\rangle}{f \cdot \langle B_I\rangle ^2}$\\
    \noalign{\smallskip}
    &&[M$_{\sun}$] & [d] & & [kG] & [kG] & [$\%$] & [kG$^2$] & [kG$^2$] & [$\%$]\\
    \noalign{\smallskip}
    \hline
    \noalign{\smallskip}
    Gl 182  &        & 0.75 & 4.35 & 0.17 & 0.17 & 2.5$^{a}\pm0.8$ & $6.3\pm1.9$ & 0.04 & \phantom{1}$ 9.1\pm3.9$ & $0.4\pm0.2$\\
    Gl 494A & DT Vir & 0.59 & 2.85 & 0.09 & 0.15 & 3.0$^{b}\pm0.8$ & $5.0\pm1.3$ & 0.03 & $11.3\pm4.3$            & $0.3\pm0.1$\\
    Gl 569A & CE Boo & 0.48 & 14.7 & 0.35 & 0.10 & 1.8$^{a}\pm0.8$ & $5.6\pm2.5$ & 0.01 & \phantom{1}$ 4.1\pm2.6$ & $0.4\pm0.2$\\ 
    Gl 388  & AD Leo & 0.42 & 2.24 & 0.05 & 0.19 & 2.9$^{c}\pm0.8$ & $6.6\pm1.8$ & 0.06 & $10.5\pm4.1$            & $0.6\pm0.2$\\
    \noalign{\smallskip}
    \hline
    \noalign{\smallskip}
    Gl 873  & EV Lac & 0.32 & 4.38 & 0.07 & 0.53 & \phantom{$>$}3.9$^{d}\pm0.8$ & \phantom{$<$}$13.6\pm2.8$ & 0.39 & \phantom{$>$}$19.0\pm5.6$ & \phantom{$<$}$2.0\pm0.6$\\
    Gl 285  & YZ Cmi & 0.31 & 2.77 & 0.04 & 0.56 & $\ga$3.9$^{c}\pm0.8$& $\la$$14.4\pm3.0$ & 0.52 & $\ga19.0\pm5.6$ & $\la2.7\pm0.8$\\ 
    \hline
    \hline
  \end{tabular}
  \flushleft
  \small
  $^{a}${this work}; $^{b}${\cite{Saar96}, see text}; $^{c}${\cite{Reiners07}}; $^{d}${\cite{JKV00}}
  \normalsize
\end{table*}

We compile measurements of mean magnetic flux from Stokes~V, $\langle
B_V \rangle$, and Stokes~I, $\langle B_I \rangle$, for the six M
dwarfs in Table\,\ref{tab:results} together with their mass,
rotational period, and Rossby number as given in \citet{Donati08} and
\citet{Morin08}. \citet{Reiners07} discuss the uncertainties of
magnetic flux measurements in FeH, they found 2-$\sigma$ uncertainties
on the order of 200\,G but estimate total uncertainties (including
systematics) on the order of a kG. The uncertainties we give for the
two new measurements in Sect.\,\ref{sect:Bfmeasurement} are 600 and
800\,G (3-$\sigma$). We adopt 800\,G as uncertainty in $\langle B_I
\rangle$ for all stars.  The ratio $\langle B_V \rangle / \langle B_I
\rangle$ is given in column~8. The fraction of magnetic flux visible
in Stokes~V is always less than 15\,\% of the total flux measured in
Stokes~I. The difference between partially convective and fully
convective stars is quite obvious: In partially convective stars
roughly 6\,\% of the total flux is detected in Stokes~V while in fully
convective stars about 14\,\% are detected.

The fraction of magnetic flux seen in Stokes~V (center panel of
Fig.\,\ref{fig:results}) never exceeds a value of $\sim$15\,\%, which
means that more than 85\,\% of the magnetic flux is invisible to
magnetic flux measurements in Stokes~V. A possible explanation is that
the vast majority of the magnetic flux on M dwarfs is organized in
small structures that are distributed over the stellar surface so that
different polarities cancel out each other in Stokes~V. This situation
would be similar to the solar case, where the strongest magnetic
fields are concentrated in spots that consist of neighbored regions of
different polarity.

Another factor to be considered are the simplifications inherent to
the results from Stokes~V used here. The technique used by
\citet{Donati08} and \citet{Morin08} makes a few important
assumptions. First, the ``weak-field approximation'' is used
\citep{Semel89}. \citet{DB97} estimate that the weak-field
approximation becomes incorrect at fields on the order of 1.2\,kG, but
they find it is still adequate to fields up to 5\,kG \cite[see
also][]{Wade00}. The weak-field approximation becomes particularly
relevant together with the second simplification, the use of a
Least-Square-Deconvolution technique \citep{Donati97, DC97, Wade00}.
In the weak-field approximation, different Land\'e factors enter the
profile as a scaling parameter that can be accounted for in a mask
(like the line-depth). This makes LSD applicable to the weak-field
case. The detectability of strong fields may be hampered because very
strong Zeeman signals cannot be fully taken into account.  It is still
an open question whether fields on the order of 2--3\,kG can be
accurately reconstructed with this method. The only way to address
this would be to compute stellar surfaces with real radiative transfer
in all lines.

For now, we can only speculate that the Stokes~V maps used for this
analysis may systematically miss magnetic fields stronger than a few
kG.  This opens a rather interesting option for the explanation of the
discontinuity at the boundary to full convection: If, in contrast to
partially convective stars, fully convective stars have magnetic flux
less concentrated in small areas of strong magnetic fields, but more
evenly distributed in large areas of weaker fields, more of the flux
may be detectable in fully convective stars.  This is an interesting
alternative that could imply a weaker degree of organization in fully
convective M dwarfs rather than a stronger large-scale component.
Unfortunately, this alternative is difficult to test with current
instrumentation. So far, the mentioned assumptions are necessary to
detect the subtle signatures of stellar magnetic fields on Stokes~V.

\subsection{Trends with mass and Rossby number}

\begin{figure*}
  \centering
  \resizebox{.7\hsize}{!}{\includegraphics{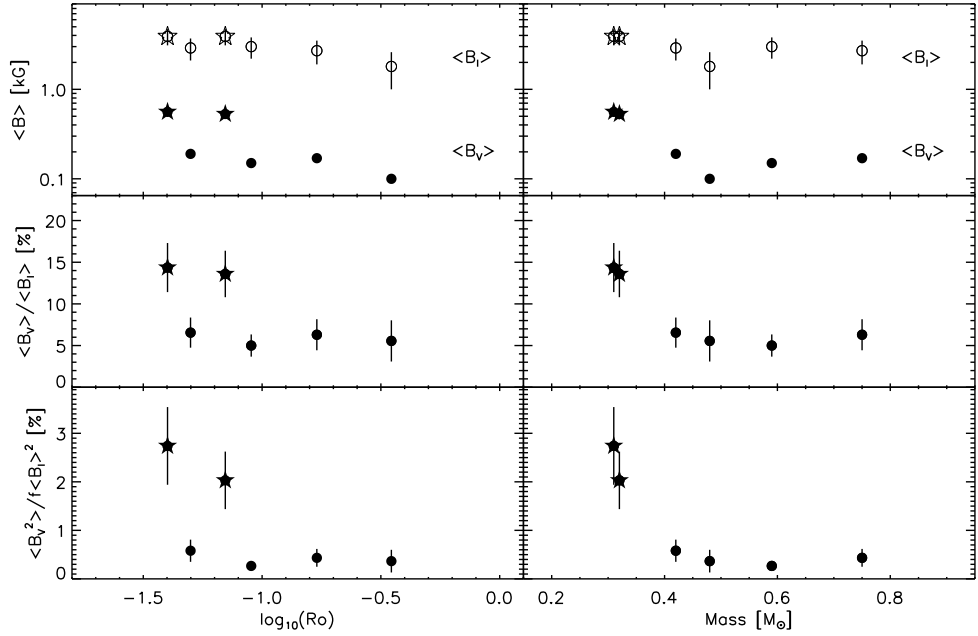}}
  \caption{\label{fig:results} \emph{Top panel:} Mean magnetic field
    measurements from Stokes~I (open symbols) and Stokes~V (filled
    symbols). \emph{Center (Bottom) panel:} Ratio of large-scale
    magnetic flux (energy) to total magnetic flux (energy). Left and
    right panels show these values as a function of Rossby number and
    mass, respectively. Symbols distinguish between fully convective
    (stars) and partially convective (circles) stars. }
\end{figure*}

The mean magnetic flux and the ratio between magnetic flux observed in
Stokes~V and Stokes~I are plotted in the upper two panels of
Fig.\,\ref{fig:results}. Left and right panels show them as a function
of Rossby number and mass, respectively. Uncertainties in $\langle B_I
\rangle$ are estimated to be 800\,G in all stars, which is mainly due
to systematic effects. No error estimates are available for $\langle
B_V \rangle$.

The upper left panel of Fig.\,\ref{fig:results} shows the
rotation-magnetic flux relation: Total magnetic flux, $\langle B_I
\rangle$, grows with smaller $Ro$ in the regime $\log{Ro} > -1$ and
saturates at lower Rossby number \citep[cp.][]{Reiners08}. Total
magnetic flux may also depend on mass with higher $\langle B_I
\rangle$ at lower masses, but mass and Rossby number to some extent
are degenerate because of the large convective overturn times in low
mass stars (at lower mass, saturation occurs at lower rotation rate).
On the other hand, the large-scale flux, $\langle B_V \rangle$, shows
a very clear dependence on stellar mass. A dependence of $\langle B_V
\rangle$ on Rossby number cannot be excluded.

The ratio between small-scale and total magnetic flux shows a
discontinuity at the boundary where stars are thought to become
completely convective: While $\langle B_V \rangle / \langle B_I
\rangle$ is about 6\,\% in early-M dwarfs with masses above
0.35\,M$_\odot$, it is about 14\,\% in mid-M dwarfs with $M <
0.35$\,M$_\odot$.

It is notoriously difficult to disentangle effects in Rossby number
and stellar mass because low-mass M dwarfs have Rossby numbers
generally lower than higher mass M dwarfs. Here, AD~Leo is
substantially more massive than EV~Lac and YZ~Cmi. On the other hand,
AD~Leo has a Rossby number lower than EV~Lac due to AD~Leo's higher
rotation rate.  Regardless of rotation, however, AD~Leo shows
substantially less small-scale magnetic flux a factor of 2--3 lower
than in EV~Lac (while its total magnetic flux is only $\sim$25\,\%
less than in EV~Lac).

\section{Magnetic energy}

Sometimes it is also interesting to investigate magnetism in terms of
magnetic energy, which is proportional to the magnetic flux squared.
From Doppler tomography, the magnetic energy measured in Stokes~V,
$\langle B_V^2 \rangle $, is reported \citep{Donati08, Morin08} and
given in column~9 of Table\,\ref{tab:results}. From Stokes~I, detailed
information about the flux distribution on the stellar surface is
usually not given so that $\langle B_I^2 \rangle $ is not available.

In order to calculate the ratio between the magnetic energies measured
in Stokes~V and Stokes~I, i.e., the ratio between the large-scale
magnetic energy to the total magnetic energy, we estimate from the
mean total magnetic flux, $\langle B_I \rangle$, the mean total
magnetic energy, $\langle B_I^2 \rangle$, which is generally not
identical to the square of the total mean flux, $\langle B_I
\rangle^2$. The difference between $\langle B_I^2 \rangle$ and
$\langle B_I \rangle^2$ is approximately equal to the Variance of the
magnetic flux distribution. In case of a completely uniform
distribution ($\sigma = 0$), the two are identical. The distribution
can be characterized by the \emph{standard deviation}, which is the
square root of the Variance, or
\begin{equation}
  \label{eq:var}
  \sigma^2 = \mathrm{Var}(B) \approx \langle B^2 \rangle - \langle B \rangle^2.  
\end{equation}

For Stokes~V, we can calculate the Variance of the magnetic flux
distribution, $\langle B_V^2 \rangle - \langle B_V \rangle^2$, and
find that $\sigma_V \sim k_V\ \langle B_V \rangle$ with $k_V$ ranging
from 0.61 to 0.83 with a mean of 0.69. From Eq.\,\ref{eq:var} we
derive $\langle B_V^2 \rangle \approx 1.5\ \langle B_V \rangle ^2$.
The distribution of magnetic flux measured in Stokes~I can be expected
to be more evenly distributed than the flux detected in Stokes~V,
because 85--95\,\% of it is not seen in Stokes~V. It is probably
distributed in rather small entities.  Thus, the Variance in magnetic
field distribution seen in $\langle B_I \rangle$ is probably smaller
than in $\langle B_V \rangle$. We make the assumption that the
Variance of the total flux distribution is $\sigma_I = 0.5\ \langle
B_I \rangle$ ($k = 0.5$ instead of 0.7 as for $\langle B_V \rangle$),
i.e., $f \approx 1.25$ in $\langle B_I^2 \rangle = f \langle B_I
\rangle ^2$. The exact choice of $f = [1.0 \dots 1.5]$ has no effect
on our results.  In the last column of Table~\ref{tab:results}, we
give the ratio between the magnetic energy detected in Stokes~V and
the estimated magnetic energy seen in Stokes~I.

The lower panel of Fig.\,\ref{fig:results} shows the behavior of
magnetic energy with Rossby number and mass. Uncertainties in the
ratio of magnetic energies include the uncertainty in the scaling
factor $f$, which is regarded as $f = 1.25 \pm 0.25$ (see above). As
expected, magnetic energy goes as magnetic flux. As a consequence of
magnetic energy being essentially the magnetic flux squared, the
discontinuity at $M \approx 0.35$\,M$_\odot$ is even more pronounced.

\section{Summary}

We compared the results of magnetic flux measurements carried out in
Stokes~V and Stokes~I in order to characterize the magnetic field
topology of early- and mid-M dwarfs. Early-M dwarfs ($<$M3.5) are
believed to harbor a radiative core while later stars are probably
fully convective. The Stokes~V results are mainly sensitive to large
scale fields because signal from magnetic areas of opposite polarity
cancel out each other. Simplifications (mainly the weak-field
approximation) may also have some influence on the detectability of
strong field components. Stokes~I is an indicator of the total mean
flux because all magnetic fields are seen regardless of polarity and
organization.

Our sample comprises four partially convective and two fully
convective stars assuming that the boundary to full convection is at
$M = 0.35$\,M$_\odot$. The ratio between magnetic flux seen in
Stokes~V and Stokes~I is around 6\,\% for partially convective stars
and 14\,\% for fully convective stars. The fraction of magnetic energy
stored in magnetic fields visible to Stokes~V is around 0.5\,\% of the
total flux in early-M dwarfs and 2.5\,\% in mid-M dwarfs. Our two main
results are the following:

\begin{enumerate}
\item In M dwarfs, more than 85\,\% (96\,\%) of the magnetic flux
  (energy) is stored in magnetic fields that are invisible to
  Stokes~V.
\item The fraction of the total magnetic flux that is detected in
  Stokes~V shows a remarkable jump at the boundary to full convection.
  In our limited sample, the fully convective stars store about 2--3
  times more magnetic flux (5 times more magnetic energy) in large
  scale fields visible to Stokes~V than partially convective M-dwarfs
  do.
\end{enumerate}

Our results partly confirm those of \citet{Donati08}; the magnetic
topologies abruptly change at the full convection boundary.
\citet{Donati08} show that the magnetic energy stored in axisymmetric
configurations is higher in fully convective stars than in partially
convective ones.  Their results were based solely on Stokes~V
measurements. In this work we showed that this trend also applies to
the fraction of large-scale to total magnetic energy. The small
fraction of flux and energy detected in Stokes~V shows the importance
of other Stokes components for a full description of a star's magnetic
topology. This continues to be a challenging task for the future.

\begin{acknowledgements}
  Some of the data presented herein were obtained at the W.M. Keck
  Observatory, which is operated as a scientific partnership among the
  California Institute of Technology, the University of California and
  the National Aeronautics and Space Administration. The Observatory
  was made possible by the generous financial support of the W.M. Keck
  Foundation.The Hobby-Eberly Telescope (HET) is a joint project of
  the University of Texas at Austin, the Pennsylvania State
  University, Stanford University, Ludwig-Maximilians-Universit\"at
  M\"unchen, and Georg-August-Universit\"at G\"ottingen. The HET is
  named in honor of its principal benefactors, William P. Hobby and
  Robert E. Eberly. We thank the referee, Michel Auriere, for a
  thorough and very constructive report. A.R.  acknowledges research
  funding from the DFG as an Emmy Noether fellow under RE 1664/4-1.
  G.B. acknowledges support from the NSF through grant AST-0606748.
\end{acknowledgements}

\end{document}